\documentclass[aps,prc,floatfix,twocolumn,footinbib,superscriptaddress]{revtex4}
\usepackage{epsfig}

\usepackage[latin1]{inputenc}

\usepackage{amssymb}
\usepackage{amsmath}
\usepackage{amsthm}

\usepackage{enumerate}
\usepackage{hyperref}
\usepackage{color}
\usepackage{graphicx}

\begin{document}
\title{Universal geometrical scaling of the elliptic flow}
\author{C. Andr\'es}
\affiliation{Departamento de F\'isica de Part\'iculas and IGFAE, Universidade de Santiago de Compostela, 15782, Santiago de Compostela, Spain}
\author{J. Dias de Deus}
\affiliation{CENTRA, Departamento de F\'isica, Instituto Superior T\'ecnico, Av. Rovisco Pais, 1049-001, Lisboa, Portugal}
\author{A. Moscoso}
\affiliation{Departamento de F\'isica de Part\'iculas and IGFAE, Universidade de Santiago de Compostela, 15782, Santiago de Compostela, Spain}
\author{C. Pajares}
\affiliation{Departamento de F\'isica de Part\'iculas and IGFAE, Universidade de Santiago de Compostela, 15782, Santiago de Compostela, Spain}
\author{Carlos A. Salgado}
\affiliation{Departamento de F\'isica de Part\'iculas and IGFAE, Universidade de Santiago de Compostela, 15782, Santiago de Compostela, Spain}

\begin{abstract}
The presence of scaling variables in  experimental observables provide very valuable indications of the  dynamics underlying a given physical process. In the last years, the search for geometric scaling, that is the presence of a scaling variable which encodes all geometrical information of the collision as well as other external quantities as the total energy, has been very active. This is motivated, in part, for being one of the genuine predictions of the Color Glass Condensate formalism for saturation of partonic densities. Here we extend these previous findings to the case of experimental data on elliptic flow. We find an excellent scaling for all centralities and energies, from RHIC to LHC, with a simple generalization of the scaling previously  found for other observables and systems. Interestingly the case of the photons, difficult to reconcile in most formalisms, nicely fit the scaling curve. We discuss the possible interpretations of this finding in terms of initial or final state effects.
\end{abstract}
\maketitle

\section{Introduction}
The discovery of a sizable elliptic flow in AA collisions, first observed at RHIC \cite{ref1,ref2} and later at LHC \cite{ref3}, turned up as an experimental major breakthrough. The observed anisotropic flow can exclusively be understood if the measured particles in the final state depend not only on the physical conditions realized locally at their production point, but also on the global geometry of the event. This non-local information can solely emerge as a collective effect, requiring strong interaction among the relevant degrees of freedom, i.e. quarks and gluons. The study of higher harmonics has also shown very interesting features, including the ridge structure seen in AA collisions \cite{ref4, ref5, ref6, ref7}, pPb collisions \cite{ref8,ref9} and also in high multiplicity pp collisions \cite{ref10}. The conventional understanding of the ridge is simply related to flow harmonics in a hydrodynamic scenario, where the description of the pPb ridge and, specially, the high multiplicity pp ridge is a challenge. 
The question is to what extent the ridge structure can be determined by the initial state effects and how these effects can be separated in the elliptic flow from the final state ones, amenable to a hydrodynamic description \cite{ref11,ref12,ref13,ref14,ref15,ref16,ref16b,ref17,ref18,ref19,ref20}. Along these lines, it is pointed out that some scaling laws satisfied by the elliptic flow can be very useful to dermine some properties of the initial stage of the collision which should be preserved by the hydrodynamic evolution \cite{ref21}. We go on with this research, showing that the experimental data on the elliptic flow of charged particles satisfy a universal scaling law related to the gluon saturation momentum. This scaling law is also satisfied by the photon data, suggesting that the elliptic flow of charged particles and photons should have a common origin. 

\section{Universal scaling law}
The experimental data for $v_2$ at RHIC and LHC energies normalized to the saturation momentum, eccentricity and radius of the collision area satisfy geometrical scaling:
\begin{equation}
\frac{v_2(p_T)}{\epsilon_1Q_s^AL} = f(\tau),
\label{eqn1}
\end{equation}
where

\begin{equation}
\epsilon_1 = \frac{2}{\pi} \int_0^{\pi/2}d\varphi \cos2\varphi\frac{R^2 - R_{\varphi}^2}{R^2}, \quad R_{\varphi}=\frac{R_{A}\sin(\varphi-\alpha)}{\sin\varphi}, 
\label{eqn2}
\end{equation}

\begin{equation}
\alpha = \arcsin(\frac{b}{2R_A}\sin\varphi),\quad R^2=\langle R_{\varphi}^2 \rangle= \frac{2}{\pi}\int_0^{\pi/2}d\varphi R_{\varphi}^2
\label{eqn3}
\end{equation}
and
\begin{equation}
\tau=\frac{p_T^2}{\left(Q_s^A\right)^2},
\label{eqn4}
\end{equation}
being $Q_s^A$ the saturation momentum, $R_A$ the radius of the nucleus and $L$ the length associated to the size of the collision area at a given impact parameter and energy. Indeed, the product $Q_s^AL$ is the inverse of the Knudsen number, i.e., the mean free path normalized to the length measured as the number of scattering centers. The scaling law (\ref{eqn1}) is tested in the range $0<\tau<1$. 

$\epsilon_1$ is a measure of the eccentricity of the collision. It does not depend on the distribution of scattering centers (partons or nucleons) in the transverse plane and it is determined only by the almond shape of the collision at a given impact parameter.

The scaling variable $\tau$ is known from the geometrical scaling verified in deep inelastic scattering, pp, pA and AA collisions \cite{ref22,ref23,ref24,ref25,ref26}, namely,
\begin{equation}
\frac{1}{N_A}\frac{dN_{ch}}{dp^2_T}= \frac{1}{Q_0^2}F(\tau).
\label{eqn5}
\end{equation}
and,

\begin{equation}
\left(Q_s^A\right)^2=\left(Q_s^p\right)^2A^{\alpha(s)/2}N_A^{1/6},
\label{eqn6}
\end{equation}
being $N_A$ the number of wounded nucleons. $\alpha(s)$ and the proton saturation momentum are given, respectively, by the equations,
\begin{equation}
\alpha(s) = \frac{1}{3}\left(1-\frac{1}{1+ \ln\left(\sqrt{s/s_0}+ 1\right) }\right)
\label{eqn7}
\end{equation}
and
\begin{equation}
\left(Q_s^p\right)^2=Q_0^2\left(\frac{W}{p_T}\right)^{\lambda},
\label{eqn8}
\end{equation}
with $Q_0 = 1$ GeV, $W = \sqrt{s} \times 10^{-3}$, $\sqrt{s_0} = 245$ GeV  and $\lambda = 0.27$.

The function $\alpha(s)$ in (\ref{eqn7}) has to do with energy conservation in the multiparticle production process. In gluon saturation models, as in the glasma picture of the color glass condensate or in string percolation, color flux tubes (strings) are formed, which subsequently give rise to particles via fragmentation. Even at moderate high energies (RHIC energies) the number of color strings is very large for central heavy ion collisions. The fragmentation of strings requires a minimum of energy, around $0.5$ GeV, to create at least a couple of hadrons. However, the total available energy is $A\sqrt{s}$ which, at low and intermadiate energies, is not enough to share with such a large number of strings. Asymptotically, the function $\alpha(s)$ goes to $1/3$ and $\left(Q_s^A\right)^{2}$ for central collisions behaves, as usual, like $A^{1/3}$. This parametrization of $\alpha(s)$ has been previously used in the framework of percolation of strings to describe the multiplicity distributions of pp and AA collisions at all centralities and rapidities and at SPS, RHIC and LHC energies \cite{ref27,ref28}. The scale $\sqrt{s_0}$ indicates when the energy-momentum conservation effects become small and the behaviour of the effective number of collisions starts to change from $N_A$ to $N_A^{4/3}$.

\section{Discussion}

\begin{figure}
\begin{minipage}[c][12.7cm][t]{.5\textwidth}
  \vspace*{\fill}
  \centering
  \includegraphics[scale=0.40]{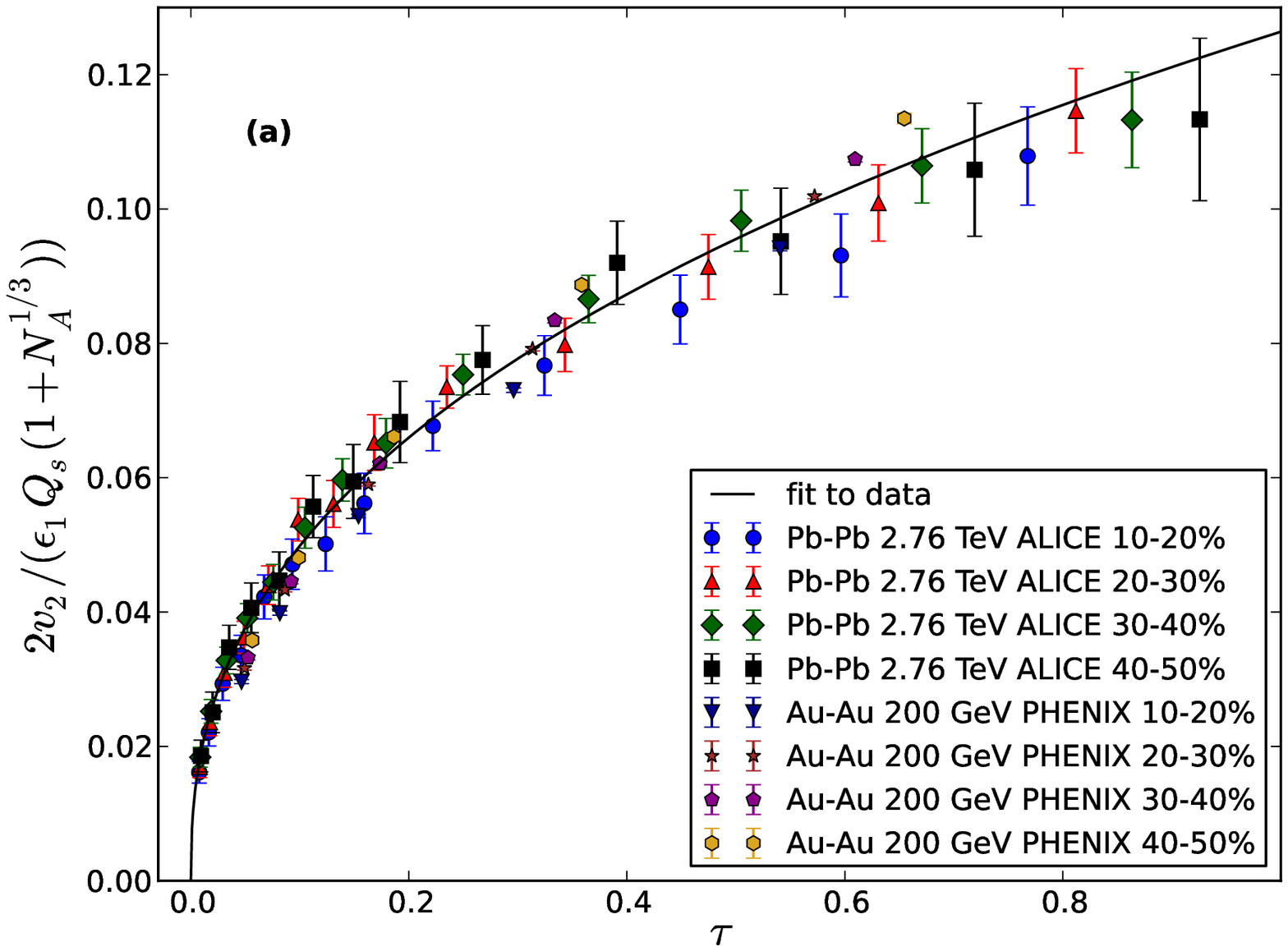}
  \includegraphics[scale=0.40]{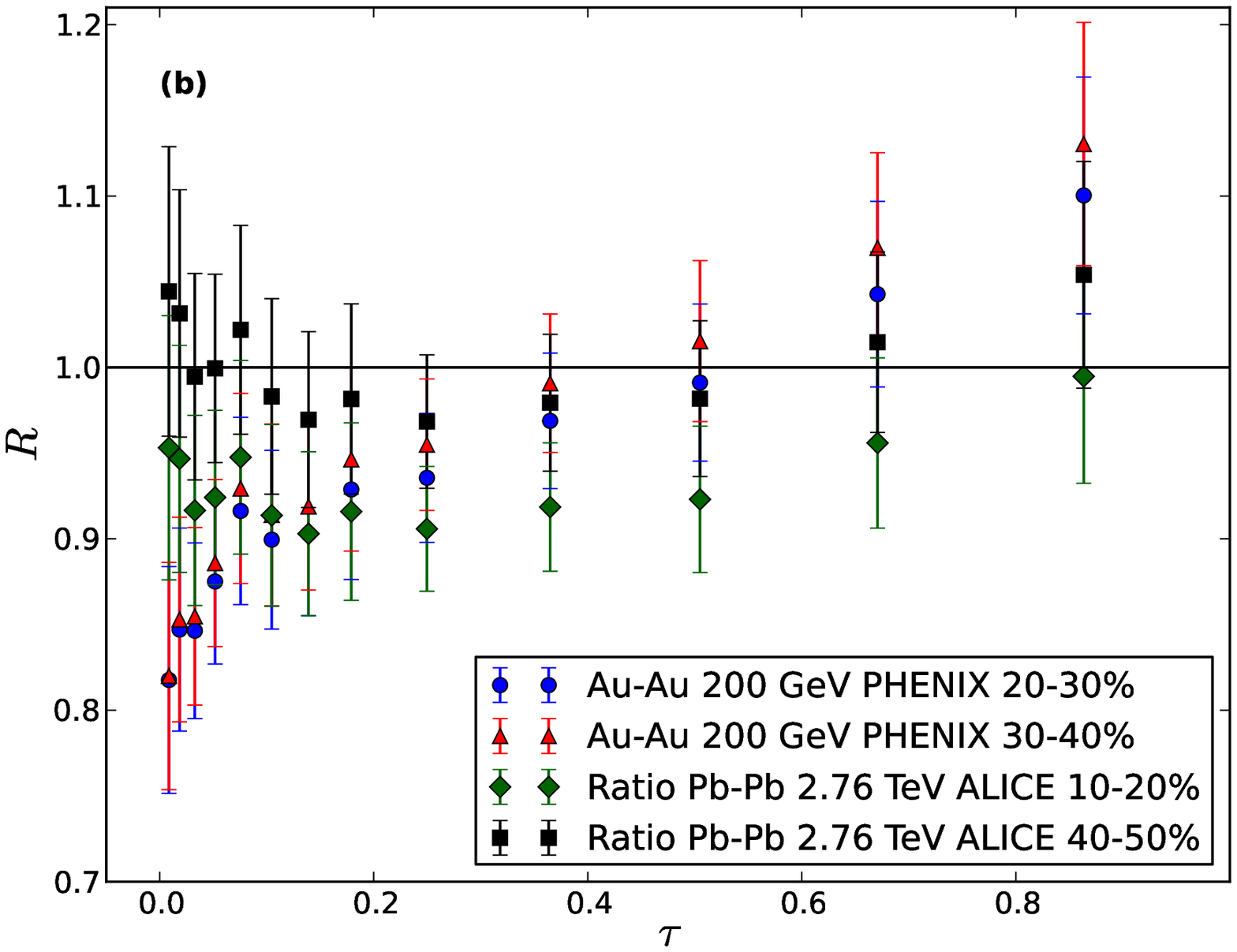}
\end{minipage}
\caption{(Color online.) \textbf{(a)} $v_2$ divided by the product $\epsilon_1Q_s^AL$ for 10-20\%, 20-30\%, 30-40\% and 40-50\% Au-Au collisions at 200 GeV \cite{ref29}, for 10-20\%, 20-30\%, 30-40\% and 40-50\% Pb-Pb collisions at 2.76 GeV \cite{ref30} in terms of $\tau$. The solid black line is a fit to data according to (\ref{eqn152}). \textbf{(b)} Ratio of Pb-Pb 10-20\%, Pb-Pb 40-50\% at 2.76 TeV \cite{ref30}, Au-Au 20-30\% and Au-Au 30-40\% at 200 GeV \cite{ref29} over Pb-Pb 30-40\% at 2.76 TeV \cite{ref30} versus $\tau$.}
\label{fig2}
\end{figure}

In Fig.~\ref{fig2} (a) we plot the measured values of $v_2(p_T)$ for Au-Au collisions for different centralities at RHIC \cite{ref29} and for PbPb collisions at LHC \cite{ref30} divided by the product $\epsilon_1Q_s^AL$ computed for each centrality and energy. We take the usual values of $b$ and $N_A$ for each centrality to compute $\epsilon_1$ and $Q_s^A$ using the equations (\ref{eqn2}), (\ref{eqn3}) and (\ref{eqn6}) respectively. $L$ is a measure of the number of longitudinal scatterings, which in the Glauber model is proportional to $N_A^{1/3}$. Nevertheless, we use $(1+ N_A^{1/3})/2$, which is used by most of the strings models as dual parton model \cite{ref31,ref32}, quark gluon string model \cite{ref33}, Venus \cite{ref34} or EPOS \cite{ref35}. In Table \ref{tab1} it is shown the corresponding values of $b$, $N_A$ and $\epsilon_1$ for each centrality and energy.
\begin{table*}[ht]
\centering
\begin{tabular}{c c c c c c c c c c c}
\hline
\hline
$\sqrt{s}$ & & \multicolumn{4}{c}{200 GeV (PHENIX)} & & \multicolumn{4}{c}{2.76 TeV (ALICE)} \\ \cline{3-6} \cline{8-11}
Centrality & & 10-20 \% & 20-30 \% & 30-40 \% & 40-50 \% & & 10-20 \% & 20-30 \% & 30-40 \% & 40-50 \% \\ 
\hline
$b\,\,(fm)$ & & 5.7 & 7.4 & 8.7 & 9.9 & & 5.6 & 7.4 & 8.9 & 10.1 \\
$N_A$ & & 117.3 & 83.3 & 57.1 & 37.2 & & 130.05 & 92.9 & 64.25 & 42.35 \\
$\epsilon_1$ & & 0.208 & 0.286 & 0.356 & 0.436 & & 0.172 & 0.238 & 0.300 & 0.357 \\
\hline
\end{tabular}
\caption{\label{tab1} Values of the impact parameter, $N_A=N_{part}/2$ and $\epsilon_1$ for PHENIX \cite{refp1} and ALICE \cite{refp2} at different centralities.}
\end{table*}
The solid black line corresponds to a fit to these data, given by
\begin{equation}
\frac{v_2}{\epsilon_1Q_s^AL}= a{\tau}^b,
\label{eqn152}
\end{equation}
where $a=0.1264 \pm 0.0076 $ and $b=0.404 \pm 0.025$. The Fig. \ref{fig2} shows that this scaling is satisfied.


In order to see the quality of this scaling we show in Fig.~\ref{fig2} (b) the ratio of Pb-Pb 10-20 \% at 2.76 TeV, Pb-Pb 40-50 \% at 2.76 TeV, Au-Au 20-30\% at 200 GeV and Au-Au 30-40\% at 200 GeV over Pb-Pb 30-40\% at 2.76 TeV as a function of $\tau$. All the ratios lie in the range $0.8-1.15$ 
for the whole $\tau$ considered, showing that the scaling is quite good (most of the experimental error data are of the order of $10$\%). 

The experimental data used in Fig. \ref{fig2} corresponds to event plane \cite{ref29} and 4-particle cumulant measurements and therefore include
some mount of fluctuations. As in the scaling law of Eq. (\ref{eqn1}) the quantities $\varepsilon_1$, $Q_s^A$ and $L$ have nothing to do with fluctuations, the mentioned fluctuations could give rise to the residual differences between the experimental data and the function $f(\tau)$ of Eq. (\ref{eqn1}). 

Changing the eccenticity, $\epsilon_1$, by the usual eccentricity, $\epsilon = \langle y^2-x^2 \rangle/\langle y^2 + x^2 \rangle$, or by the participant eccentricity, the scaling is not satisfied for both Monte-Carlo Glauber and Color Glass distributions. This fact does not mean that the initial state should give the corresponding eccentricity of a hard profile, such as it is defined in Eq. \ref{eqn2}. Probably, the scaling law could be preserved using other eccentricities, but in this case somes changes in the dependence of $L$ and $N_A$ are necessary.

\begin{figure}
\begin{minipage}[c][11.7cm][t]{.5\textwidth}
  \vspace*{\fill}
  \centering
  \includegraphics[scale=0.58]{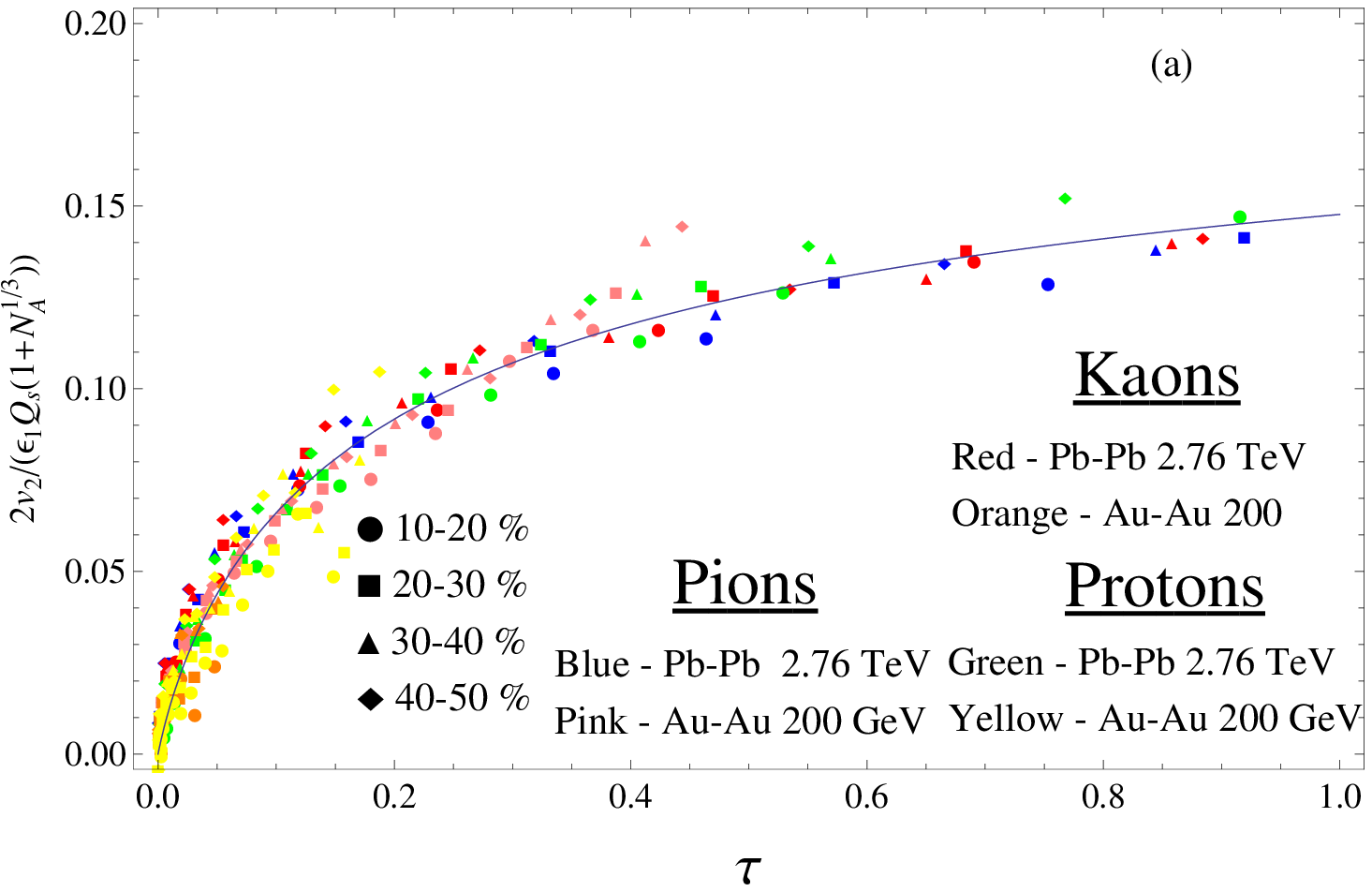}
  \includegraphics[scale=0.58]{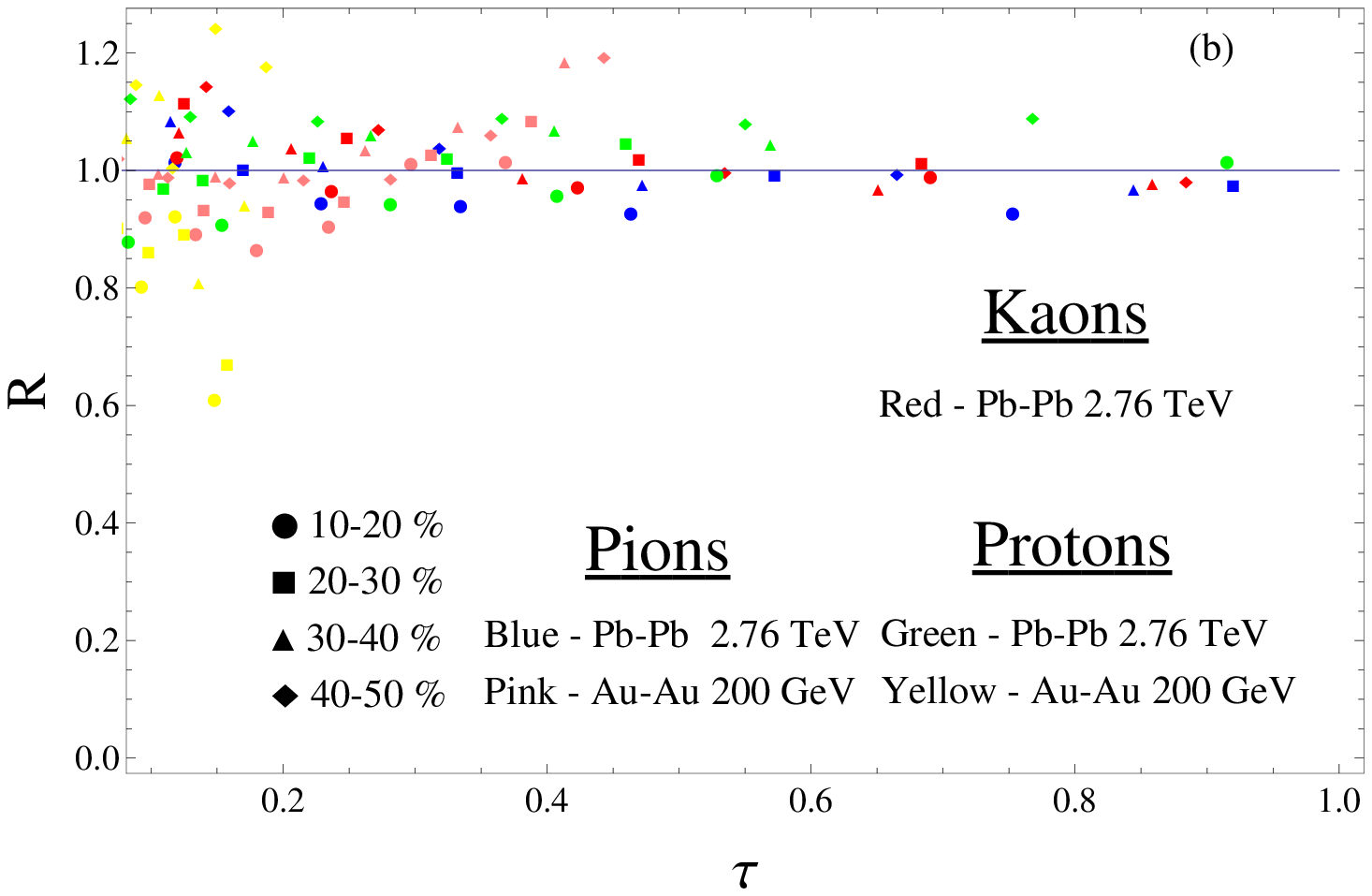}
\end{minipage}
\caption{(Color online.) \textbf{(a)} $v_2$ of $\pi$, $k$ and $p$ divided by the product $\epsilon_1Q_s^AL$ for 10-20\%, 20-30\%, 30-40\% and 40-50\% Au-Au collisions at 200 GeV \cite{ref41}, for 10-20\%, 20-30\%, 30-40\% and 40-50\% Pb-Pb collisions at 2.76 GeV \cite{ref42} versus $\tau$, for $\tau<1$. The solid line is a fit to Eq. (\ref{eqnp}) \textbf{(b)} Ratio of the experimental points over the fitting function Eq. (\ref{eqnp}) versus $\tau$, for $0.1<\tau<1$.
}
\label{fig0}
\end{figure}

The scaling law is also satisfied for specified particles, such as $\pi$, $k$ and $p$ as it is seen in Fig.~\ref{fig0}. In the case of protons, we have used an effective transverse momentum $Q_s^{'A}$ instead of $Q_s^A$, $\left(Q_s^{'A}\right)^{2} = N_s^{0.045}\left(Q_s^A\right)^{2}$, where $N_s$ is the number of strings. It is known that for central collisions the ratio baryon/meson increases with $p_T$ up to a moderate value of transverse momentum. In central collisions, due to the strong color field formed, the color flux tubes in the glasma picture or the cluster of strings in the string percolation approach have a larger string tension, producing high mass particles more efficiently. In addition to that, inside a cluster of many strings the flavour of each single string recombines with the flavour other strings, producing also baryons more eficiently. Concerning the $p_T$-distributions, these two effects can be taken into account in an effective way, defining a $Q_s^{'A}$ for baryon production, related to $Q_s^A$ by a factor which was obtained \cite{ref35b} by fitting the dependence on the number of collisions, and therefore in the number of strings, $N_s$, of the ratio of the $p_T$ integrated nucleon distribution over the $p_T$ integrated pion distribution. For this study the data provided by PHENIX was used \cite{refmm}, resulting in a $N_s^{0.09}$ dependence which in terms of $Q_s^2$ is $N_s^{0.045}$ \cite{refn}.  We observe that all the data lie in the same curve, parametrized by
\begin{equation}
\frac{v_2}{\epsilon_1Q_s^AL}= \frac{\tau}{a+b\tau+c\sqrt{\tau}}\,,
\label{eqnp}
\end{equation}
with $a=0.573\pm0.011$, $b=4.76\pm0.23$ and $c=1.52\pm0.34$.
Although in this case the scaling is not as perfect as the obtained for charged particles, we obtain a good agreement for $\tau>0.1$, with discrepancies not higher than 20\% for most of the experimental points. For $\tau<0.1$ a great departure occurs, probably motivated by the precision of our fit in this low $\tau$ region and the proximity with $\Lambda_{QCD}$. Points in this region are not shown in Figure \ref{fig0} $(\mathbf{b})$ in order to keep a good visibility of the remaining points to evaluate the scaling law. 

\begin{figure}
\includegraphics[scale=0.40]{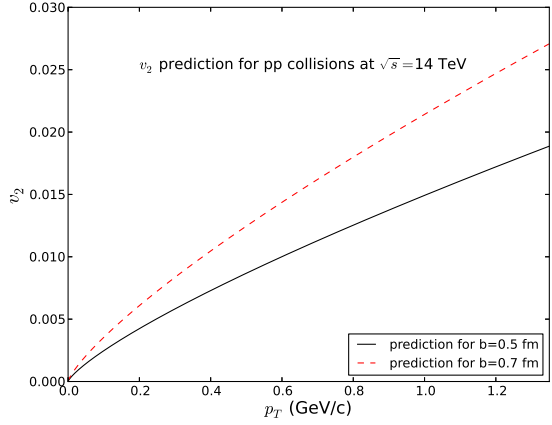}
\caption{(Color online.) $v_2$ prediction for pp collisions at $\sqrt{s} = 14$ TeV for impact parameters values of $b= 0.5$ fm (solid black curve) and $b = 0.7$ fm (dashed red curve) as a function of $p_T$.
}
\label{fig3}
\end{figure}

Assuming that the $v_2$ scaling can be extended to pp collisions, we compute the elliptic flow as a function of the trasnsverse momentum, $v_2(p_T)$, for $\tau < 1$. In Fig.~\ref{fig3} we show our predictions for $\sqrt{s} = 14$ TeV and for impact parameters values of $b= 0.5$ fm and $b = 0.7$ fm.The $v_2(p_T)$ obtained is much smaller than the computed one using hot spots inside the proton \cite{ref36} and only slightly smaller than the one found considering usual impact parameter distributions \cite{ref37,ref38}. For $b = 0.7$ the multiplicity  would not be very different from the minimum bias which in many models is predicted to be around $7.2$ at central rapidity \cite{ref27}. The relation between $b$ and the multiplicity is obtained in string like models, computing the number of strings formed in the collision assuming a particular profile function for the proton. An example of this kind of recent evaluation can be found in \cite{refu}.

\begin{figure}
\includegraphics[scale=0.40]{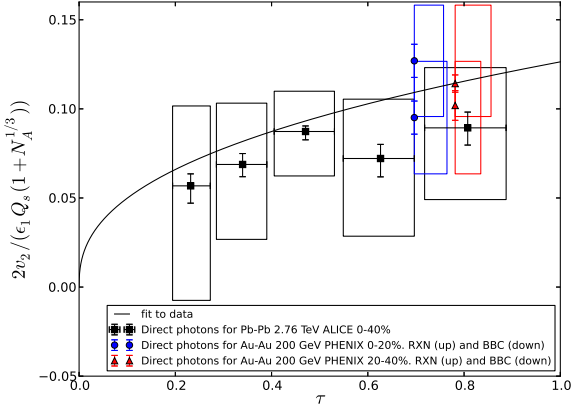}
\caption{(Color online.) $v_2$ divided by the product $\epsilon_1Q_s^AL$ for direct photons at 0-20\% and 20-40\% Au-Au collisions at 200 GeV \cite{ref40} and direct photons at 0-40\% Pb-Pb collisions at 2.76 GeV \cite{ref39} plotted as a function of $\tau$. The solid black line is the scaling curve of Figure (\ref{fig2}). We have included all the data with $p_T<Q_s$.
}
\label{fig4}
\end{figure}

We have not included in our analysis the $v_2$ data on pPb collisions due to the uncertainties in the values of $N_A$ at a given impact parameter.

In addition, since direct photon production satisfies geometrical scaling \cite{ref26}, its elliptic flow may be of the same size and $p_T$ shape of the rest of particles. In order to check this point, in Fig.~\ref{fig4} we plot the ALICE preliminary data \cite{ref39} and the PHENIX data \cite{ref40} at different centralities. PHENIX collaboration quote two different points at the same $p_T$ and centrality obtained by different analysis methods (BBC and RXN detectors). In any case, we observe that the data are close to the scaling curve.

In order to relate the geometrical saling of the transverse momentum distribution, equation (\ref{eqn5}), with the scaling of the elliptic flow of equation (\ref{eqn1}), we define an azimuthal angle saturation momentum, $Q_{s\varphi}^{A}$. As, on average, the gluon density is larger for smaller $R_{\varphi}$ (smaller $\varphi$), we assume $\left(Q_{s\varphi}^{A}\right)^2 \sim 1/R^2_{\varphi}$. In addition,  $Q_{s\varphi}^{A}$ should be proportional to the inverse of the mean free path, $\lambda_{mfp}$, normalized to the length size of the scattering, $L$, i.e. inversely proportional to the Knudsen number $k_n = \frac{\lambda_{mfp}}{L}=\frac{1}{Q_s^AL}$. Therefore, we can write:
\begin{equation}
\left(Q_{s\varphi}^{A}\right)^2 \approx \frac{L}{\lambda_{mfp}}\frac{1}{R^2_{\varphi}} = \frac{1}{k_nR^2_{\varphi}} = \frac{Q_s^AL}{R^2_{\varphi}}.
\label{eqn9}
\end{equation}

Notice that $\langle \left( Q_{s\varphi}^{A}\right)^2\rangle$ is not $\left(Q_{s}^{A}\right)^2$, indeed,

\begin{equation}
\langle \left(Q_{s\varphi}^{A}\right)^2\rangle \equiv \frac{Q_{s}^{A}L}{\langle R_{\varphi}^2 \rangle}=\frac{Q_{s}^{A}L}{R^2}\simeq \frac{\lambda_{mfp}\left(Q_{s}^{A}\right)^2}{L} \simeq k_n\left(Q_{s}^{A}\right)^2\,,
\label{eqn9b}
\end{equation}
where we do the approximations $\langle 1/R_{\varphi}^2\rangle \approx 1/\langle R_{\varphi}^2\rangle$ (the values for LHC of $\langle 1/R_{\varphi}^2\rangle$ are 0.0353, 0.0365, 0.0522 and 0.0769 for the centralities 10-20\%, 20-30\%, 30-40\% and 40-50\%, respectively. The corresponding values of $1/\langle R_{\varphi}^2\rangle$ are 0.0334, 0.0344, 0.0467 and 0.0644. Similar values hold for RHIC data) and $L\approx R$. The last is a rough approximation which we use for simplicity. It just intends to stablish the order of magnitude of $L$. 

As $k_n = \frac{\lambda_{mfp}}{L}$ is very small for heavy nuclei collisions, $\langle \left(Q_{s\varphi}^{A}\right)^2\rangle <<\left(Q_{s}^{A}\right)^2$.

Now, the scaling variable, $p_T^2/\left(Q_{s}^{A}\right)^2$, should be replaced by
\begin{equation}
\tau_{\varphi}=p_T^2\left(\frac{1}{\left(Q_{s}^{A}\right)^2}+\frac{1}{\left(Q_{s\varphi}^{A}\right)^2}\right),
\label{eqn9c}
\end{equation}

which encodes the azimuthal dependence. After integrating this $\varphi$-dependence, the geometrical scaling (\ref{eqn5}) of the transverse momentum distribution should be recovered. $Q_{s\varphi}$ is responsible of the scaling of the elliptic flow $(Q_{s}^{A}$ is independent of the azimuthal angle and therefore does not contribute to the elliptic flow) but its contribution to the $p_T$ distribution is negligible. Actually, inserting Eq. (\ref{eqn9c}) into Eq. (\ref{eqn5}) we obtain the distribution $F(\tau_\varphi)$. Since the transverse momentum distribution $F(\tau)$ is a strongly decreasing function (exponentially) of $\tau$ , $F(\tau_\varphi)$ decreases as well. As $\langle \left(Q_{s\varphi}^{A}\right)^2\rangle <<\left(Q_{s}^{A}\right)^2$, $\exp(-p_T^2/Q_{s\varphi}^2)$ decreases faster than $\exp(-p_T^2/(Q_{s}^{A})^2)$ and its contribution to the
azimuthal integrated distribution is negligible.

Therefore, elliptic flow reads
\begin{equation}
v_2 = \frac{4\int_0^{\pi/2} d\varphi \cos2\varphi\frac{dN}{dp_T^2d\varphi}}{\frac{dN}{dp_T^2}}=\frac{4\int_0^{\pi/2} d\varphi \cos2\varphi F(\tau_{\varphi})}{F(\tau)}.
\label{eqn10}
\end{equation}

Expanding $F(\tau_{\varphi})$ in powers of $R_{\varphi}^2-R^2$ and retaining the first non-vanishing term, we have,
\begin{equation}
v_2 = \frac{2}{\pi}\int_0^{\pi/2} d\varphi \cos2\varphi \frac{R^2 -R_{\varphi}^2}{R^2}\frac{4}{ F(\tau)}\tau Q_s^AL\left. \frac{dF}{d\tau_{\varphi}}\right|_{R_{\varphi}^2=R^2},
\label{eqn12}
\end{equation}
where we assume that the spatial angle $\varphi$ is now the azimuthal angle of the emitted particle. This assumption encodes the different density that the outgoing particle had to probe in its path along the almond. In string percolation models, a higher density means a higher multiplicity. Therefore, with a $\varphi$-dependent density (or, equivalently, a $\varphi$-dependent $Q_s$), the angular distribution $\frac{dN}{dp_T^2d\varphi}$ must be higher for angles with higher density. This is exactly the case if we use the aforementioned assumption. We also approximate, again, $L \approx R$.

For $R_{\varphi}^2=R^2$, we have $\tau_{\varphi}=\tau\left(1+Q_s^AL\right)$. If $\left. \frac{dF}{d\tau_{\varphi}}\right|_{R_{\varphi}^2=R^2}$ is a decreasing function of $\tau_{\varphi}$, the additional dependence $Q_s^AL\tau$ would give a negligible contribution for $\tau \lesssim 1/Q_s^AL$. This means that for $\tau < 0.2-0.6$ there will be sizable contributions spoiling the scaling. Indeed, the $v_2$ computed directly from $dN/dp_T^2d\varphi$ assuming the equation (\ref{eqn10}), is very different from the experimental data for $\tau < 0.5$. In Eq. (\ref{eqn10}) we identify the spatial angle with the angle of the momentum of the emitted particle.

Although the scaling law can not be derived from the geometrical scaling in a simple way, it would be interesting to know the origin of this scaling and the role of saturation in it. In the affirmative case, it would be also interesting to look for a scaling law similiar to (\ref{eqn1}) for the rest of the harmonic moments. The total distribution might factorize in two terms: one with the product of the number of scatterings and a scaling function on $p_T^2/\left(Q_s^A\right)^2$ and the other with the sum of the products of the different eccentricities with the corresponding azimuthal dependence.

\section{Conclusions}

It is shown that the experimental data on the eliptic flow of charged particles for Au-Au and Pb-Pb collisions for different centralities at RHIC and LHC energies satisfy a scaling law. The elliptic flow for identified particles: $\pi$, $k$ and $p$ lies in the same curve. The photon data, despite their large uncertainties, also satisfy this scaling. Other than the eccentricity, this scaling law involves the number of scatterings and a function which depends only on $p_T^2/\left(Q_s^A\right)^2$. The number of scatterings in the only involved quantity in relation with final state effects. The rest has to do with the geometry and gluon saturation.

\section*{Acknowledgements}

We thank N. Armesto for very useful discussions. This work has been done under the project FPA2011-22776 of MINECO (Spain), the Spanish Consolider CPAN project, FEDER funds and Xunta de Galicia.

\bibliographystyle{apsrev}
\bibliography{v2}
\end{document}